\documentclass[%
 reprint,
 amsmath,amssymb,aps,secnumarabic, graphics,floatfix,nofootinbib,
 tightenlines,nobibnotes,aps,prl,12pt
]{revtex4-1}
\usepackage{mathrsfs}
\usepackage{amsfonts} 
\usepackage{amssymb} 
\usepackage{float} 
\usepackage{graphicx}
\usepackage{dcolumn}
\usepackage{bm}
\usepackage[font=small,justification=justified,format=plain]{caption}
\usepackage{subcaption}
\DeclareCaptionJustification{justified}{\justifying}
\usepackage[colorlinks,citecolor=blue,linkcolor=blue,urlcolor=blue,bookmarks=false,hypertexnames=true]{hyperref}

\begin{document}

\title{Calculation of tunneling current across Trapezoidal potential barrier in a Scanning Tunneling Microscope}
\thanks{Corresponding author. Arun V. Kulkarni Tel.:0832 2580309; fax: +91-8322557033 E-mail address: avkbits@goa.bits-pilani.ac.in}%

\author{Malati Dessai}
 \altaffiliation[Also at ]{
 	Department of Physics, Parvatibai Chowgule College of Arts and Science, Margao, Goa, 403602.}
\author{Arun V Kulkarni}%
 \email{avkbits@goa.bits-pilani.ac.in}
\affiliation{%
 	Department of Physics, Birla Institute of Technology and Science-Pilani,K K Birla Goa campus, Goa, 403726.\\
 }%
\date{\today}
\begin{abstract}
The Planar Model of the Electrode-Vacuum-Electrode configuration for STM in which electrode surfaces are assumed to be infinite parallel planes, with atomic size separation and vacuum between them, is used to calculate tunneling current densities for both low and high bias voltages. Non WKB, Airy function solutions for the Schr\"odinger Equation for the trapezoidal barrier in the tunneling region are used to calculate the tunneling probability. Temperature dependent Fermi Factors for each electrode are introduced and the calculation involves integration over the electron energies. In order to convert the current densities obtained in the planar model to tunneling currents the tip and sample surfaces are modelled as confocal hyperboloids, and the tip sample distance is replaced by the length of the line of force (field line). The current is found by integrating the current density over a finite area of the tip. The calculated tunnel currents for a few electrode pairs at room temperature are plotted for several values of bias voltage, tip sample distances, and tip radii of curvature. Pauli Effects are studied as a function of bias voltage and tip-sample distance. Some estimate of lateral resolution and its dependence on bias voltage and tip radius is also presented.
 

\end{abstract}
\date{\today}
\maketitle
\section{\label{sec:level1}Introduction}
Scanning Tunneling Microscopes (STM's) allow imaging of surface structures with atomic scale resolution and have been the workhorses for most surface science studies in recent decades. Scanning probe microscopes usually involve a sharp tip brought close to a sample which is mostly flat except for atomic scale surface features. The atoms of the tip interact with those of the sample and in case of the STM this interaction causes tunelling of the electrons between the tip and the sample. An experimentally measurable quantity such as the tunneling current is measured as a function of the tip sample distance and the bias potential between tip and sample.  In typical STM applications, extremely sharp tips are used and further a very close proximity between the tip and sample is maintained.  As a result, the tip-sample interaction is highly localized and the tunneling current is sensitive only to the local surface properties of the sample. This sensitivity is further enhanced due to the STM current having an almost exponential dependence on the tip-sample distance, thus achieving atomic scale resolution\cite{binnig5}. Since the tunneling is a very sensitive function of the tip sample distance, the measured values of the tunnel current almost resemble the sample surface topography when raster scanning of the sample is carried out by the tip.
\par There are several other works in literature, that calculate tunneling current density such as Simmons\cite{simmonsI}, Hartman and Chivian\cite{HartmanChi} in which the WKB approximation is used for calculating the tunneling probability. While tunneling current densities are not directly experimentally measurable  tunneling currents are directly measurable and these are of particular interest in comparing theory with experiment. 
\par Tersoff and Hamann\cite{TersoffHamann} have developed a formalism starting from Bardeen's theory\cite{BardeenI}\cite{BardeenII} to calculate tunneling currents. However, their calculations is likely to work best for very low tip-sample distances and also for very low biases.
In this paper the tip-sample distances exceed this limit and range from a few $\AA$ to 20 $\AA$. The calculation of Chen et.al.\cite{chen} uses Bessel functions of one-third order and therefore goes beyond the WKB approximation. However only the current densities are reported and effects of Pauli Blocking are not investigated. The self consistent calculation of Banerjee et.al.\cite{SBPZ} uses the WKB approximation. They have not studied Pauli Blocking effects and also report only current densities and not currents.
\par The tip and sample here are initially treated as plane surfaces. In this so called planar model, only tunneling current densities and not tunneling currents can be calculated. The latter would be infinity in a planar model. In the non planar model considered here, the tip and the sample are modelled as confocal surfaces in a prolate spheroidal coordinate system. In this paper the tunneling current densities are converted into currents through a method described by Saenz and Garcia\cite{SG}. 
\par The energies of electrons inside the electrodes are distributed according to the Fermi-Dirac distribution. With these assumptions the tunneling current density and tunneling currents through the barrier are calculated for both high and low bias voltages.

\section{\label{sec:level2}THEORY}
\subsection{\label{sec:level5}A. CURRENT DENSITY AND FERMI FACTOR}
Consider two plane conducting electrodes with infinite plane surfaces, separated by vacuum whose width ranges from  2 \AA\ to 20 \AA.  The electrode surfaces are initially assumed to be of infinite area. With this assumption, both the Laplace equation for the electrostatic potential, and the Schr\"odinger equation with this potential, reduce to one dimensional equations.
\begin{figure}[t]
	\centering
	\begin{subfigure}[b]{0.5\textwidth}
		\centering
		\includegraphics[width=3.0in,height=2.2in]{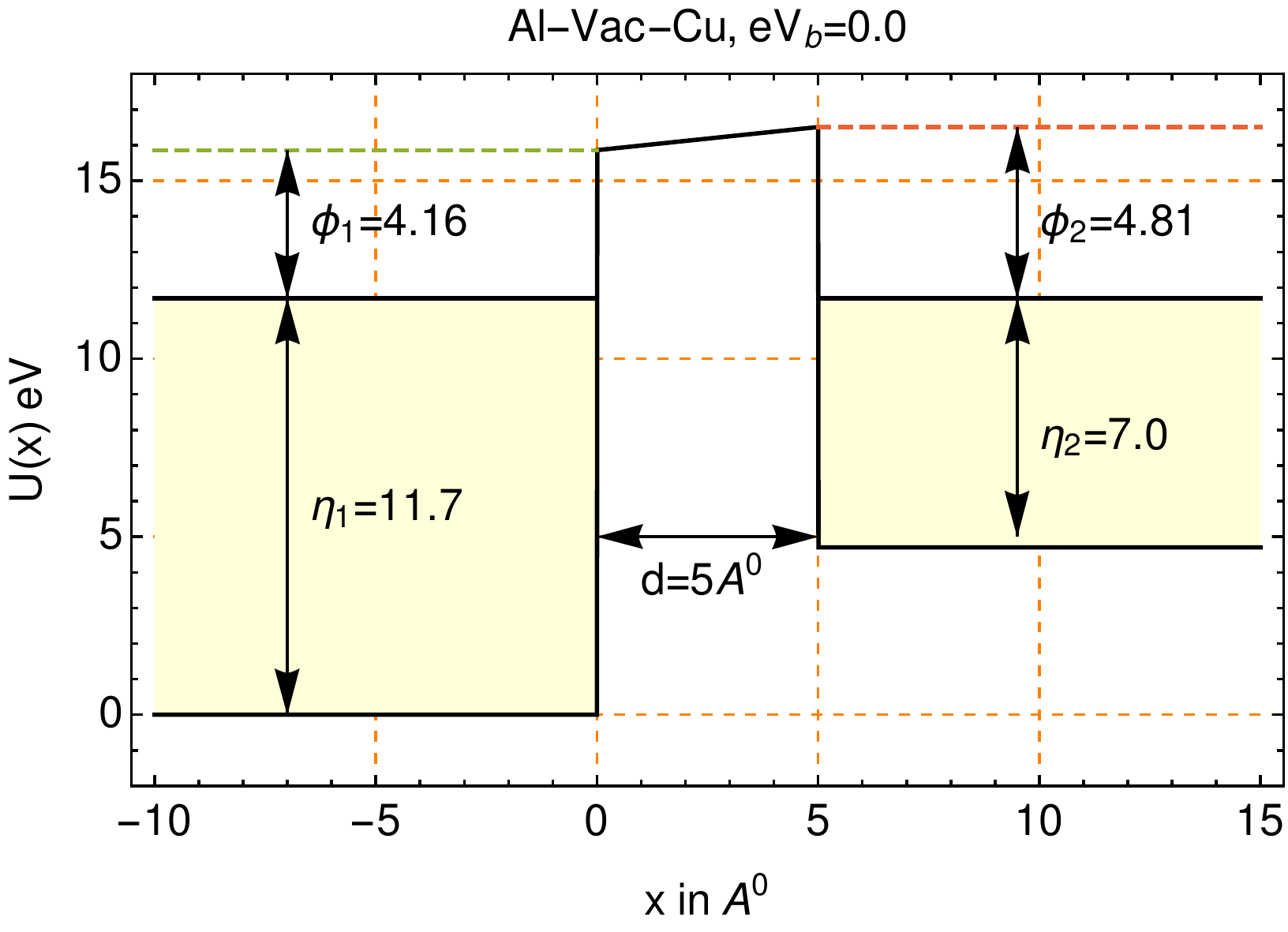}
		\caption{zero bias}			
		\label{fig:1a}
	\end{subfigure}	
	\begin{subfigure}[b]{0.5\textwidth}
		\centering
		\includegraphics[width=3.0in,height=2.2in]{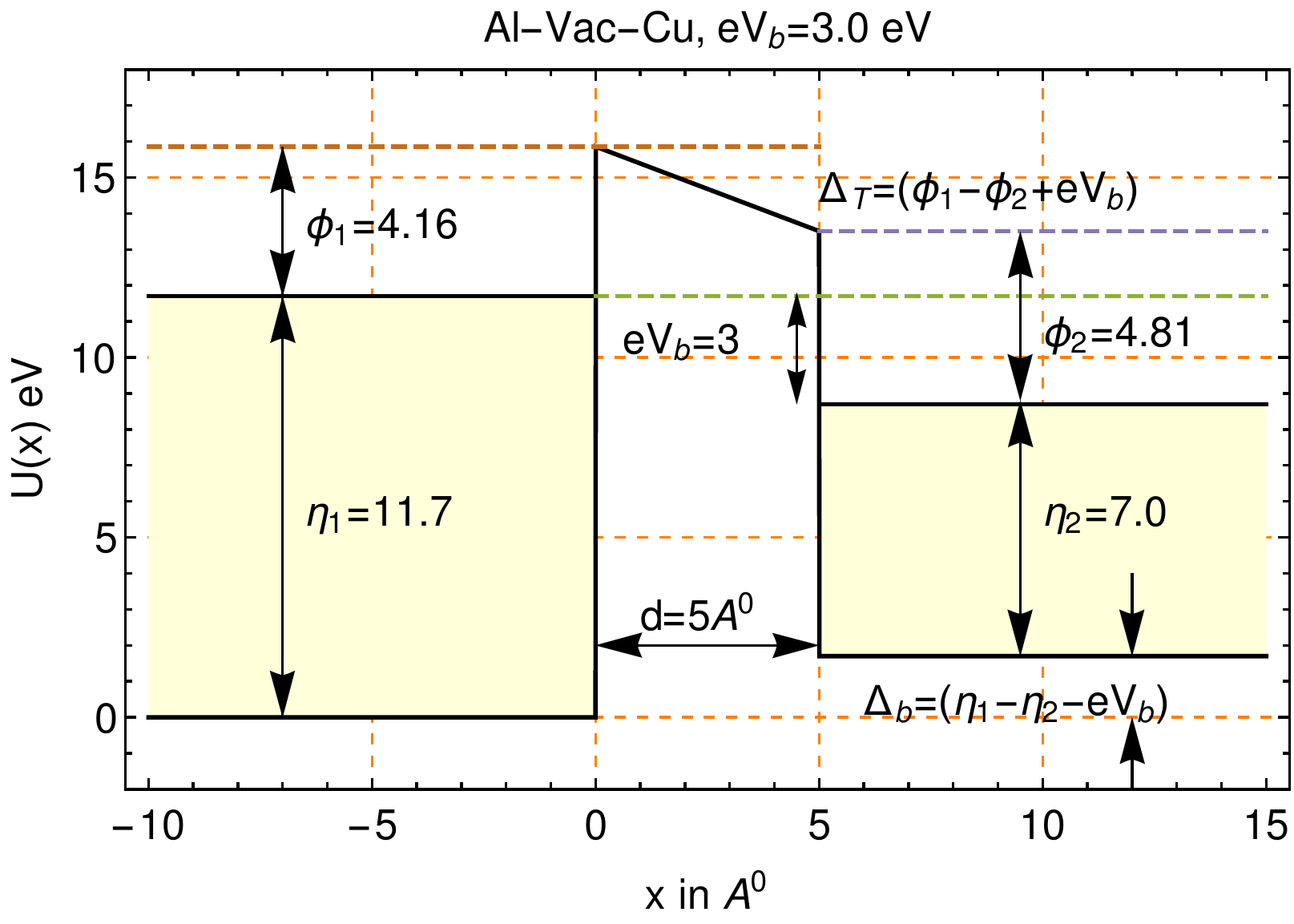}
		\caption{$\text {Bias}$ = 3.0 eV $ (eV_b < \phi_1)$}
		\label{fig:1b}
	\end{subfigure}	
	\caption{Dissimilar Electrodes with vacuum barrier at a) Zero Bias b) Bias = 3.0 eV }	
	\label{fig:Figure0}	
\end{figure}
\par
\begin{figure}[h]
	\centering
	\includegraphics[width=3.0in,height=2.3in]{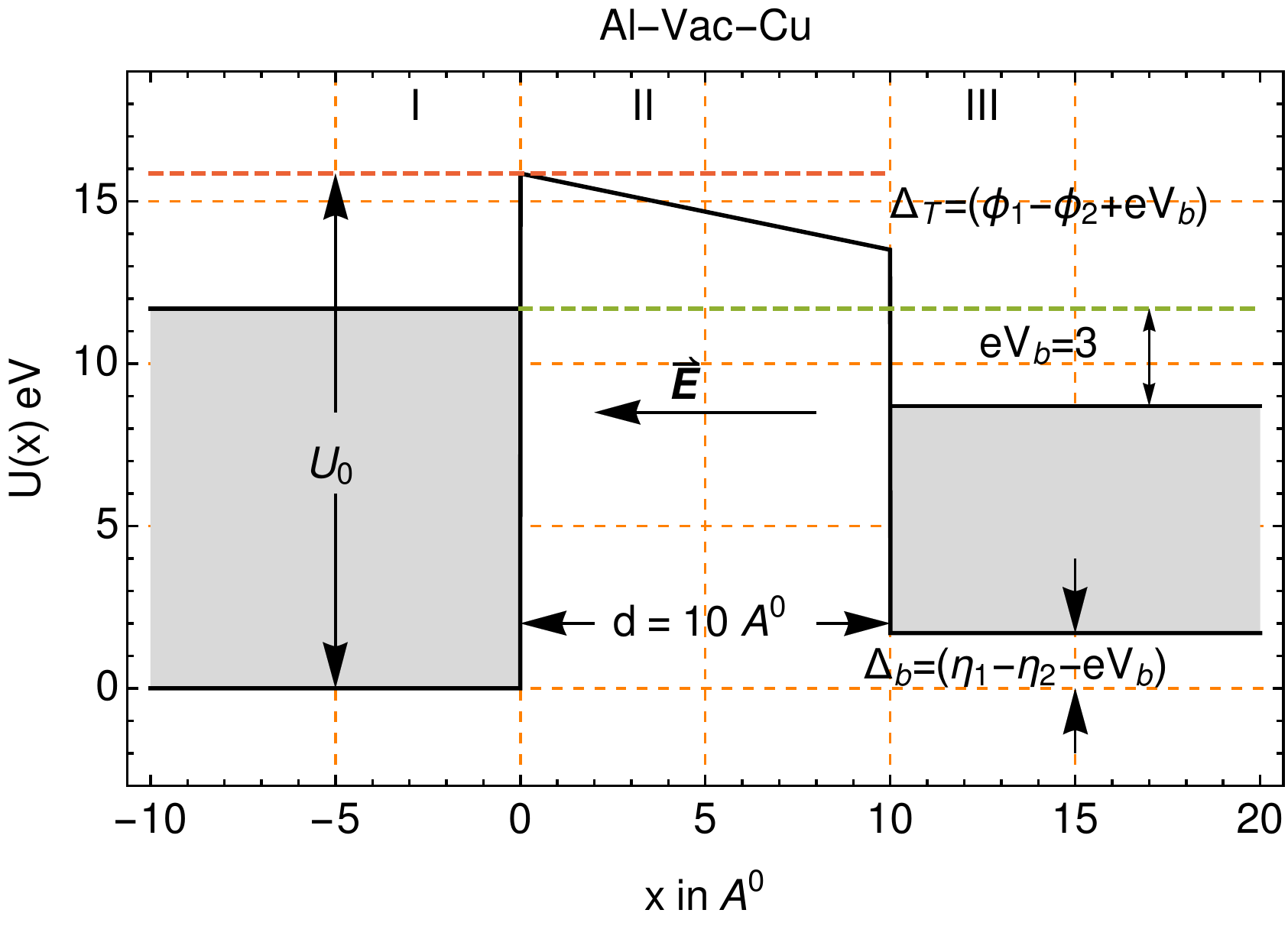}		
	\caption{Electrode 1 is at higher potential ($U_0=\eta_1+\phi_1$) and the elctric field points along the negative x axis}
	\label{fig:FigureA}
\end{figure}

 Let a bias potential $V_b$ be applied across the electrodes. Let $\eta_1$, $\phi_1$, $\eta_2$, $\phi_2$, be the Fermi energies and work functions of the left (first) electrode and the right (second) electrode respectively (Fig.\ref{fig:1a}). The left electrode (electrode 1) is assumed to be at higher potential than the right electrode (electrode 2) so that the electric field points along the negative x axis, and if a free electron were introduced between electrodes it would experience a force along the positive x axis as shown in Fig.\ref{fig:FigureA}. Tunneling of electrons between the two plane surfaces occurs through the effective trapezoidal barrier in the vacuum region. The case for zero bias between dissimilar electrodes is shown in Fig.\ref{fig:1a}. The barrier potential in this case is the difference $(\phi_1-\phi_2)$ between the work functions of the two conducting electrodes.  When biased, the barrier potential in the vacuum region, is the sum of the contact potential $(\phi_1-\phi_2)$ and the applied external bias $eV_b$ as shown in Fig.\ref{fig:1b}. Forward electron current density is  along the positive x axis and reverse electron current density is along the negative x axis.
\par The connection between the tunneling current density and the probability of tuneling through the barrier, is well established and the earliest accounts of it are by Fowler and Nordheim\cite{FN} and Simmons\cite{simmonsI}\cite{simmonsII}. In these accounts, one of the principal assumptions is that the number of electrons that can occupy a unit volume in the 6 dimensional phase space is $\dfrac{2}{h^3}$, and therefore the number per unit volume (of coordinate space) of electrons in the velocity space volume element $d^3 v = dv_x \,dv_y \,dv_z$ about the velocity vector ${\bf v}$ is $\mathfrak{n}(v)d^3 v =\frac{2m}{h^3} d^3 v$
The energy of these electrons is 
\begin{eqnarray}\nonumber
E = \frac{1}{2} m {\bf v}^2 & = & \frac{1}{2} m( v_x^2 +v_y^2 +v_z^2)\\
& = & E_x+E_y+E_z
\end{eqnarray}
 where $E_x=\frac{1}{2}m v_x^2, E_y=\frac{1}{2}m v_y^2, E_z=\frac{1}{2}m v_z^2$ are the kinetic energies of the electrons for translational motion along the $x, \, y, \, \text{and} \, z$ directions respectively. The number density of electrons in the electrode 1 that are available for tunneling would be $n_1 \, d^3 v = \mathfrak{n}(v) f_1(E) \, d^3 v$ where $f_1(E) = [1 + \exp({\beta(E-\eta_1)})]^{-1}$ is the Fermi Dirac Distribution function which can be regarded as the probability of occupation of the energy level $E$ in the electrode 1. Note $\beta = \frac{1}{kT}$ where $k$ is the Boltzmann Constant and $T$ is the temperature (in Kelvin) of both the electrodes.
\par The number density of electrons that can tunnel through the barrier is $$n_T(v) d^3 v = n_1(v) D(E_x) \,d^3 v$$ where $D(E_x)$ is the probability of tunneling through the barrier. In theoretical calculation $$D(E_x)= \rvert T(E_x) \rvert^2 \frac{j_{out}}{j_{inc}}$$ where $T$ is the transmission amplitude for tunneling through the barrier and if plane wave solutions are used within the electrodes then ${j_{inc}}=\frac{\hbar k_1}{m}$ and ${j_{out}}=\frac{\hbar k_2}{m}$ for forward currents. where $\hbar k_1$ and $\hbar k_2$ are x-components of the momenta of the electrons in the first and the second electrodes.
\begin{equation}
\hbar k_1=\sqrt{2mE_x}, \, \,  \hbar k_2= \sqrt{2m(E_x-\Delta_B)} 
\end{equation} 
where $E_x$ is the kinetic energy of motion of the electrons in the $x$- direction. For reverse currents ${j_{inc}}=\frac{\hbar k_2}{m}$ and ${j_{out}}=\frac{\hbar k_1}{m}$.
Note that the transmission amplitude is independent of the direction of tunneling. Hence it is the same for both forward and reverse tunneling currents.
Not all of the electrons emitted by electrode 1 will necessarily be accepted by the electrode 2, since there will be partial occupation of this level by electrons already present in the electrode 2. This is due to the Pauli Exclusion Principle and may be called the Pauli Effect. Others\cite{simmonsI}\cite{Shu7}\cite{riza}\cite{Ebeling} have referred to this phenomena as Pauli Blocking.
\par To obtain the number density of electrons in the energy level $E$, that are accepted by the electrode 2, one must multiply $n_T$  by the probability of vacancy of this energy level, in the electrode 2. This is given by $ [1 - f_2(E+\Delta_b)]$ where $$f_2(E+\Delta_b) = [1 + \exp({\beta(E+\Delta_b-\eta_2)})]^{-1}$$ 
The energy argument in the Fermi Dirac function for electrode 2 will be shifted by $+\Delta_b=\eta_1-\eta_2-eV_b$, because the zero level of the energy in electrode 2 is shifted by $-\Delta_b$ compared to that of the electrode 1. But as can be seen from Figures  \ref{fig:FigureA} and \ref{fig:1c} 
$$(E+\Delta_b-\eta_2)=(E+eV_b-\eta_1)$$ Therefore 
$$f_2(E+\Delta_b)=f_1(E+eV_b)$$
\par Thus the number density of electrons that will be found in the states of electrode 2, and which contribute to the forward current are 
\begin{align}\nonumber
& n_\text{For} d^3 v\\&= \dfrac{2m^3}{h^3}\frac{k_2}{k_1}\lvert T(E_x) \rvert^2 f_1(E)[1 - f_2(E+\Delta_b)] d^3 v \\&=\dfrac{2m^3}{h^3}\frac{k_2}{k_1}\lvert T(E_x) \rvert^2 f_1(E)[1 - f_1(E')] d^3 v
\end{align}
where $ E'=E+eV_b=E_x+E_r+eV_b$ 
The forward current density is then $$J_\text{For} = e\int v_x\, n_\text{For}\, d^3 v$$ Now $v_x dv_x = (dE_x/m)$ and $$dv_y dv_z = v_r dv_r d\theta = (dE_r/m) d \theta$$ where the $y, z $ velocity components are expressed in terms of their polar components $(v_r, \theta)$ in the $y- z$ plane. Also $E_r = \frac{1}{2} m v_r ^2 $  and $E = E_x + E_r$. Putting it all together and integrating over $\theta$ gives
\begin{multline}
$$ J_\text{For} =\dfrac{4\pi me}{h^3} \int dE_x \dfrac{k_2}{k_1}\lvert T(E_x) \rvert^2\,\times\\ \int\limits_0^\infty dE_r f_1(E)[1 - f_1(E')]$$
\end{multline} 

In a like manner one can also calculate the reverse current density to be 
\begin{multline}
$$J_\text{Rev} = \dfrac{4\pi me}{h^3} \int dE_x \dfrac{k_1}{k_2}\lvert T(E_x) \rvert^2\,\times\\ \int\limits_0^\infty dE_r [1 - f_1(E)] f_1(E') $$
\end{multline}
The net current density is given by $J_\text{Net} = J_\text{For} - J_\text{Rev}$, which becomes 

\begin{multline}
$$J_\text{Net} = \dfrac{4\pi me}{h^3} \int dE_x \lvert T(E_x) \rvert^2\,\times\\ \int\limits_0^\infty dE_r \Big [f_1(E)[1 - f_1(E')]\,\dfrac{k_2}{k_1} -\\ [1 - f_1(E)] f_1(E')]\,\dfrac{k_1}{k_2} \Big] $$
\end{multline}
Define the Fermi Factor $\mathcal{F}(E_x)$ as
\begin{multline}\label{np}
$$\mathcal{F}(E_x)=\int\limits_0^\infty dE_r \Big [f_1(E)[1 - f_1(E')] \,\dfrac{k_2}{k_1} -\\ [1 - f_1(E)] f_1(E')\Big]\, \dfrac{k_1}{k_2} $$
\end{multline}
Carrying out the integral over $E_r$
\begin{multline}\label{pb}
 $$\mathcal{F}(E_x)= \dfrac{k_2}{k_1}F_1(E_x) - \dfrac{k_1}{k_2}F_1(E_x+eV_b)\, + \\ \frac{\Big [\dfrac{k_1}{k_2}-\dfrac{k_2}{k_1}\Big ]}{(1-e^{-\beta eV_b})}\Big [F_1(E_x+eV_b)-e^{-\beta eV_b}F_1(E_x)\Big] $$ 
\end{multline}
where $$F_1(E_x)=\dfrac{1}{\beta}\text{Log}[1+e^{-\beta(E_x-\eta_1)}]$$ 
\begin{figure}[h]
	\centering
	\includegraphics[width=3.4in,height=3.5in]{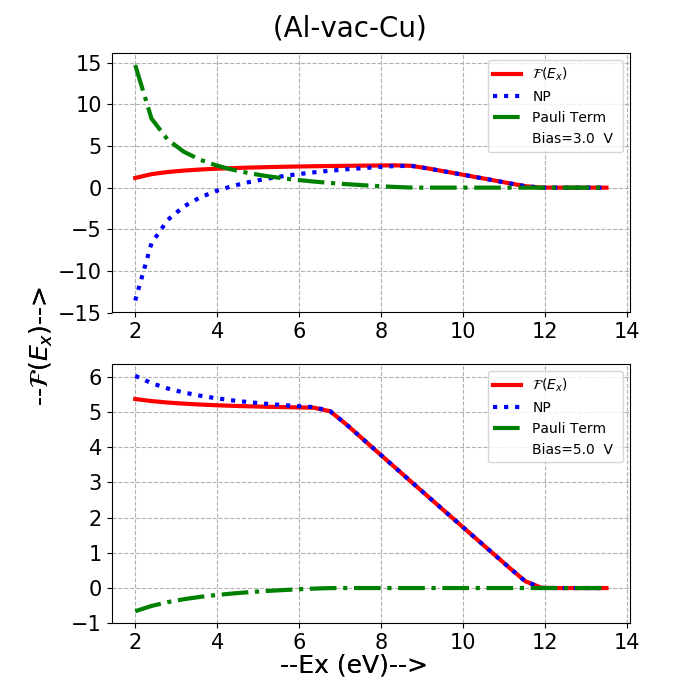}	
	\caption{Fermi Factor for Bias voltage of $3$ and  $5$ V for Al-vac-Cu. The solid red line refers to $\mathcal{F}(E_x)$, \,the \,dotted \,\,line\,\, refers\, to \,\,Non-Pauli term, and dot-dashed line refers to Pauli term.}
	\label{fig:FF}
\end{figure}

\begin{figure}[h]
	\centering
	\includegraphics[width=3.0in,height=2.1in]{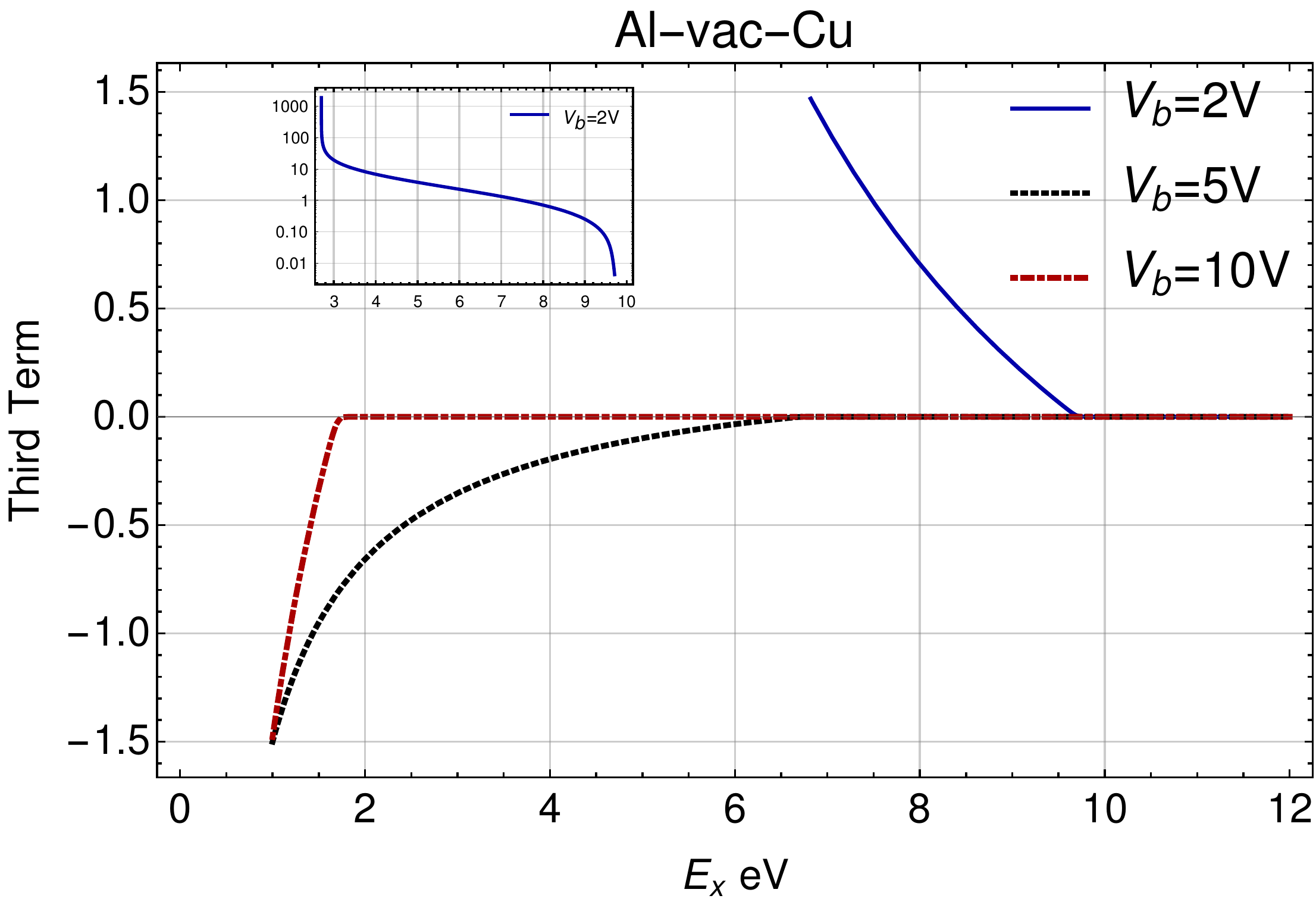}	
	\caption{Third Term as a function of $E_x$ for bias voltages $2$V,$5$V,and $10$V for Al-vac-Cu.}
	\label{fig:ThTerm1}
\end{figure}

\begin{figure}[h]
	\centering
	\includegraphics[width=3.0in,height=2.1in]{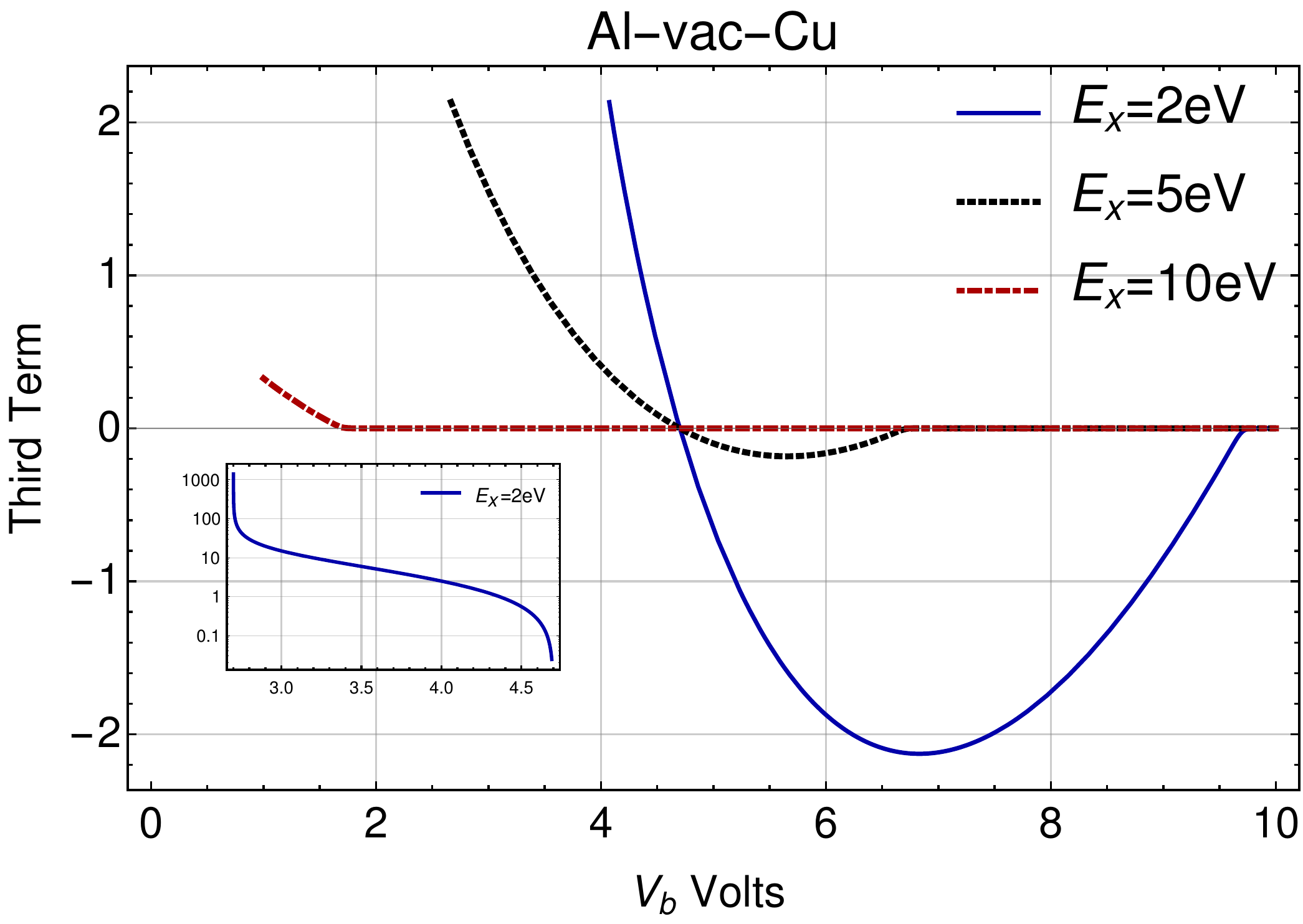}	
	\caption{Third Term as a function of $V_b$ for electron energies of  $2$eV,$5$eV,and $10$eV for Al-vac-Cu.}
	\label{fig:ThTerm2}
\end{figure}

 The first two terms are the Non-Pauli contribution; these are the terms that one would get if Pauli Blocking were completely ignored. The third term in equation (\ref{pb}) explicitely introduces Pauli Effects. The behaviour of these terms $viz$ Non-Pauli, $\mathcal{F}(E_x)$, and the Pauli term (third term only), as a function of $E_x$ is displayed in Fig.\ref{fig:FF}.   
The Fermi Function $\mathcal{F}(E_x)$ (solid red line) is seen to peak at $ 8.7 \,\text{eV}$ $\approx \eta_1-eV_b$ for $eV_b=3 \,\text{eV}$; and it peaks at $ 6.7\, \text{eV}$ $\approx \eta_1-eV_b$ for $eV_b=5  \, \text{eV}$. Subsequently  $\mathcal{F}(E_x)$ is found to decrease rapidly to zero as $E_x$ exceeds $\eta_1$. This is found true for all bias voltages. The dotted line refers to the Non-Pauli(NP) term, which may have negative values for low bias voltages. For these voltages the reverse current exceeds the forward current. This may be due to the different flux ratios for the forward and the reverse currents. The dot-dashed line refers to the third term of equation (\ref{pb}). The third term is seen to go to zero at about $E_x=\eta_1-eV_b$. This term contains Pauli Effects and it is studied in more detail in Figures \ref{fig:ThTerm1} and \ref{fig:ThTerm2}.
\par Figure \ref{fig:ThTerm1} plots the third term for fixed bias voltages ($V_b=2$V, $V_b=5$V,$V_b=10$V) as a function of electron energies in the range ($1$ eV to $12$ eV). The graph shows that Pauli Effects are significant for low energy electrons and especially for low bias voltages. For $E_x > \eta_1-eV_b$ the third term is essentially zero implying that there is negligible Pauli Effect. The inset in this figure shows the third term for $2$V rising rapidly by at least $4$ orders of magnitude for very low values of $E_x$. Figure \ref{fig:ThTerm2} plots the third term for fixed electron energies ($E_x=2$eV, $E_x=5$eV,$E_x=10$eV) as a function of the bias voltage in the range ( $1$V to $10$V). The inset in this figure shows the third term for $E_x=2$eV rising rapidly by at least $3$ orders of magnitude for very low values of bias voltage. All curves in Fig.\ref{fig:ThTerm2} pass through zero at $eV_b=\eta_1-\eta_2$ which in case of electrodes Al and Cu is given by $V_b = 4.7$ V. This is because $k_1 = k_2$ for this value of bias, and the zero level of the energy in the two electrodes become identical. For bias voltages lesser than $eV_b=\eta_1-\eta_2$ the third term is positive and its effect is to increase the net tunneling current beyond the non-Pauli contribution. For bias voltages larger than this value the third term becomes negative and the Pauli Effect will be to reduce the net tunneling current from its Non-Pauli value.
\par The net current density becomes
\begin{equation}\label{jnet}
J_\text{Net} = \dfrac{4 \pi m e}{h^3} \int\limits_{E^{\text{min}}_x} ^{E^{\text{max}}_x} dE_x  \lvert T(E_x) \rvert^2 \mathcal{F}(E_x)
\end{equation} 
where the limits of the integral in equation (\ref{jnet}) are given by 
$$E_x^\text{min}=\text{Max}[0,\Delta_b]\, \,\text{and}\,\, E_x^\text{max}=\eta_1+\phi_1$$
$E_x^\text{min}$ is the minimum of energy in the energy Stage I and $E_x^\text{max}$ is the maximum of energy in the energy Stage II. The energy stages I, II, and III are described in the next section. 

\subsection{\label{sec:levelA}B. TUNNEL AMPLITUDE ${\bm {T(E_x)}}$}
\begin{figure}[h]
	\centering
	\includegraphics[width=3.0in,height=2.2in]{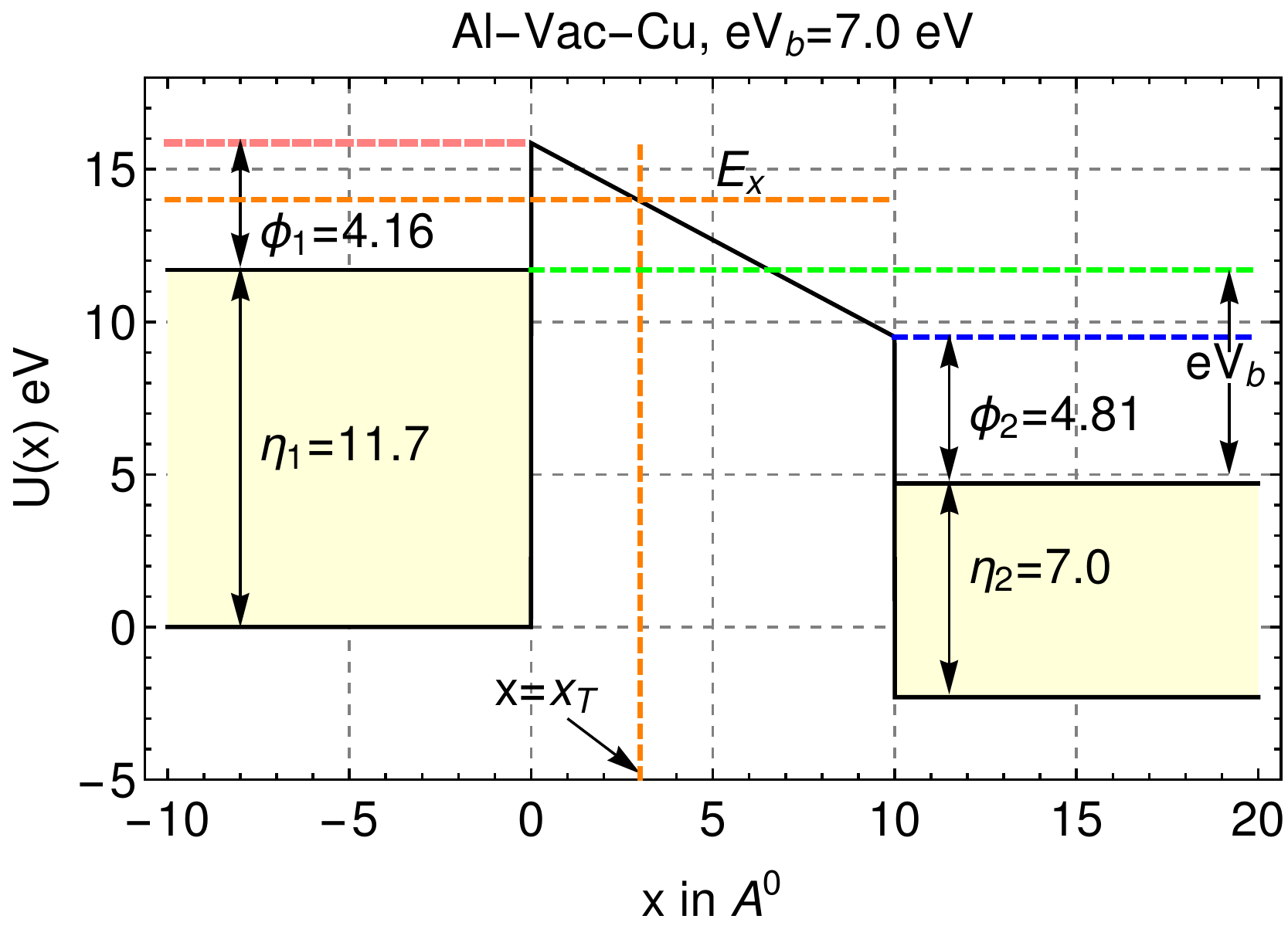}
	\caption{Energy Stages I ($0$ eV to $9.51$ eV), II ($9.51$ eV to $15.86$ eV), and III ($> 15.86$ eV) for Al-vac-Cu, and $eV_b$  = 7.0 eV $ (eV_b > \phi_1)$. The upper limits are indicated by blue dotted line (Stage I), pink dotted line (Stage II).}
	\label{fig:1c}
\end{figure}

Let the two planar electrodes described in the previous section be parallel to each other and parallel to the y-z plane. Let their surfaces be at $x=0$ and $x=d$ respectively, with the intervening space being occupied by vacuum. The entire extent of the x-axis can be divided up into three spatial regions and the potential energy of an electron in these three regions is given by   
\begin{equation}
	U(x)=\left\{
	\begin{array}{@{}ll@{}}
	0 & 		x < 0 \\
	U_{II}(x) & 		0 \leqslant x \leqslant d \\
	\Delta_b&	 x > d\\
	\end{array}\right.
\end{equation} \\
	where
\begin{equation}
	U_{II}(x)=(\eta_1+\phi_1)-(\phi_1-\phi_2+eV_b) \dfrac{x}{d}
\end{equation} 	
The wavefunctions in two regions $(x < 0)$ and $(x > d)$ are given by   
\begin{equation}\label{psi1}
	\psi_1(x)=e^{ik_1x}+Re^{-ik_1x}, \hspace*{.2in} x < 0		
\end{equation}
\begin{equation}\label{psi3}
\psi_3(x)=Te^{ik_2x}, \hspace*{.2in} x > d
\end{equation}
where, $R$, $T$ are the reflection and transmission amplitudes.
The wavefunction $\psi_b$ in the barrier region $0 \leqslant x \leqslant d$ is given by the  Schr\"odinger equation for the linear (trapezoidal) potential
\begin{equation}
\frac{d^2 \Psi_b}{dx^2}-(A-Bx)\Psi_b= 0
\end{equation} 
where
\begin{equation}
	 A=\dfrac{2m}{\hbar^2} (\eta_1+\phi_1-E_x) 
\end{equation}	
\begin{equation}
	B=\dfrac{2m}{\hbar^2 d} (\phi_1-\phi_2+eV_b) 
\end{equation} 
If the electrodes were at absolute zero temperature, only electrons at the Fermi levels of the electrodes would undergo tunneling. At higher temperatures, electrons with a wider range of energies will be involved in tunneling. The energy range of such electrons can be divided into three Energy Stages. Fig.\ref{fig:1c} shows the three Energy Stages for Al-vac-Cu system for which $\eta_1=11.7$ eV, $\phi_1=4.16$ eV,$\eta_2=7.0$ eV, $\phi_2=4.81$ eV, $V_b = 7$ volts.\\

\noindent\textbf{Energy Stage I}: The barrier region in this stage does not contain a turning point. The energy in this region ranges from Max$[0,\Delta_b]<E_x<(\eta_1+\phi_1-eV_b)$. Therefore, the term $(A-Bx)$ will never be zero $\forall \, x \, \in \, [0,d]$. \\
\noindent\textbf{Energy Stage II}: In this stage, there is a turning point inside the barrier region at $x = x_T = \frac{A}{B}$. The energy range in this stage is $(\eta_1+\phi_2-eV_b)<E_x<(\eta_1+\phi_1)$ as shown in Fig.\ref{fig:1c}. The barrier is triangular and electrons tunneling out of this triangular barrier are real (not virtual) electrons which have not entered the the second electrode and those that do so enter the second electrode, contribute to the tunneling current. \\
\noindent\textbf{Energy Stage III}: The energy of the electrons in this stage exceeds $(\eta_1+\phi_1)$ and these electrons should be emitted without having to tunnel through any barrier suggesting that the transmission probability would be very high. However for most values of $\phi_1$ the electron energy is so far above the Fermi energy $\eta_1$ that the Fermi factor (for room temperatures) is essentially zero and the contribution of electrons in this energy stage at these temperatures to the current density is vanishingly small. Therefore unless the electrode temperatures are very high (i.e. comparable to the Fermi temperature of the electrodes), the contribution of electrons in this Energy Stage will be neglected alltogether.\\
\par
\begin{figure}[h]
	\centering
	\includegraphics[width=3.5in,height=2.5in]{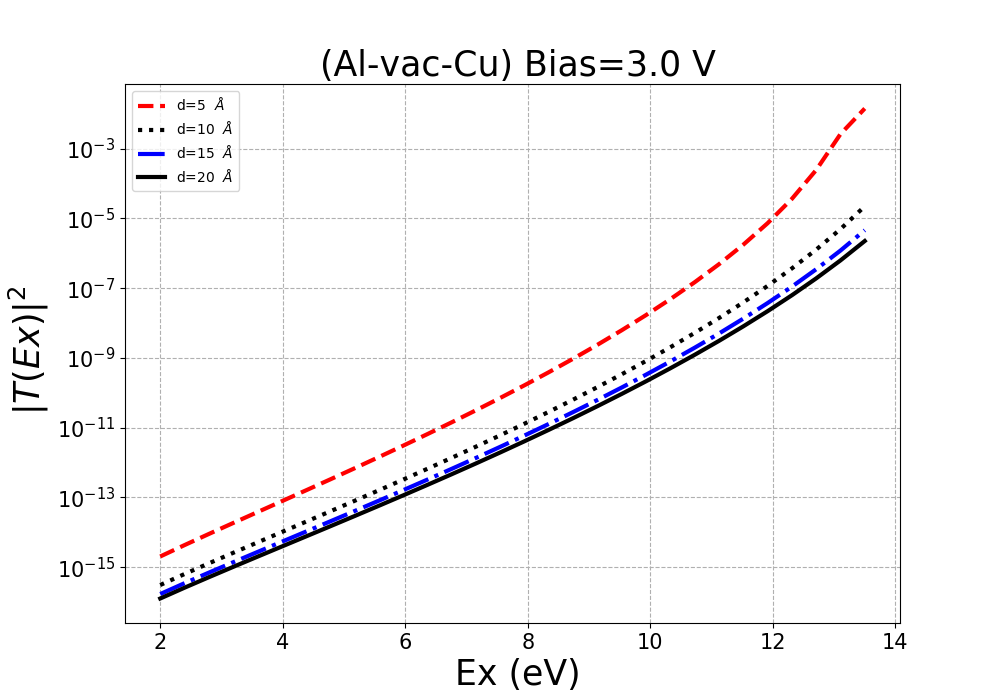}	
	\caption{Plot of Transmission Probability versus $E_x$ for Bias Voltage = 3 V $\&$ $d=5,10,15,20$ $\AA$ }
	\label{fig:DEx}	
\end{figure}

 In Energy Stages I and II the Schr\"odinger equation (4) reduces to the Airy equation.
\begin{equation}
\frac{d^2 \Psi_b}{dh^2} - h\Psi_b= 0
\end{equation} 
where $h=h(x)=\dfrac{A}{B^{2/3}}-B^{1/3}x$.
The general solution of the Airy Differential equation can be written as \cite{AS}
\begin{equation}\label{phden18}
\psi_b(x)= C \phi^{(1)}(x)+ D \phi^{(2)}(x)
\end{equation} 
where
\begin{equation}\label{phden19}
\phi^{(1)}(x)= \dfrac{Ai[h(x)]}{Ai[h(\bar{x})]}\,\,
\phi^{(2)}(x)= \dfrac{Bi[h(x)]}{Bi[h(\bar{x})]}
\end{equation}
The denominator of the functions $\phi^{(1)}$ and $\phi^{(2)}$ contain Airy functions evaluated at $\bar{x}$ (a fixed value of x near the centre of the barrier region). These denominators are sometimes required to avoid overflow and underflow errors in the calculation of the wavefunctions for different values of $x$ in the barrier. The constants $C$ and $D$ are determined from matching of the wavefunction and its derivative at the relevant spatial boundaries. In Energy Stage I these boundaries are at $x=0$ and $x=d$. In Energy Stage II, two separate wavefunctions $\psi_{b1}$ and $\psi_{b2} $ have to be determined for $x < x_T$, and $ x > x_T$ respectively. The spatial boundaries for the former are at $x=0$ and $x=x_T$ and that for the latter, are at $x=x_T$ and $x=d$. The Tunneling amplitude is given by 
\begin{equation}\label{eq8}
	T(E_x)=e^{-i k_2 d}\Big\{	
\begin{array}{@{}ll@{}}
 \psi_b(d)  & \text{in Stage I} \\
 \psi_{b2}(d) & \text{in Stage II} 
\end{array}
\end{equation} 
\noindent Fig.\ref{fig:DEx} shows that the $\rvert T(E_x) \rvert^2$ increases with $E_x$ which is to be expected.
\par
\begin{figure}[h]
	\centering
	\includegraphics[width=3.5in,height=2.5in]{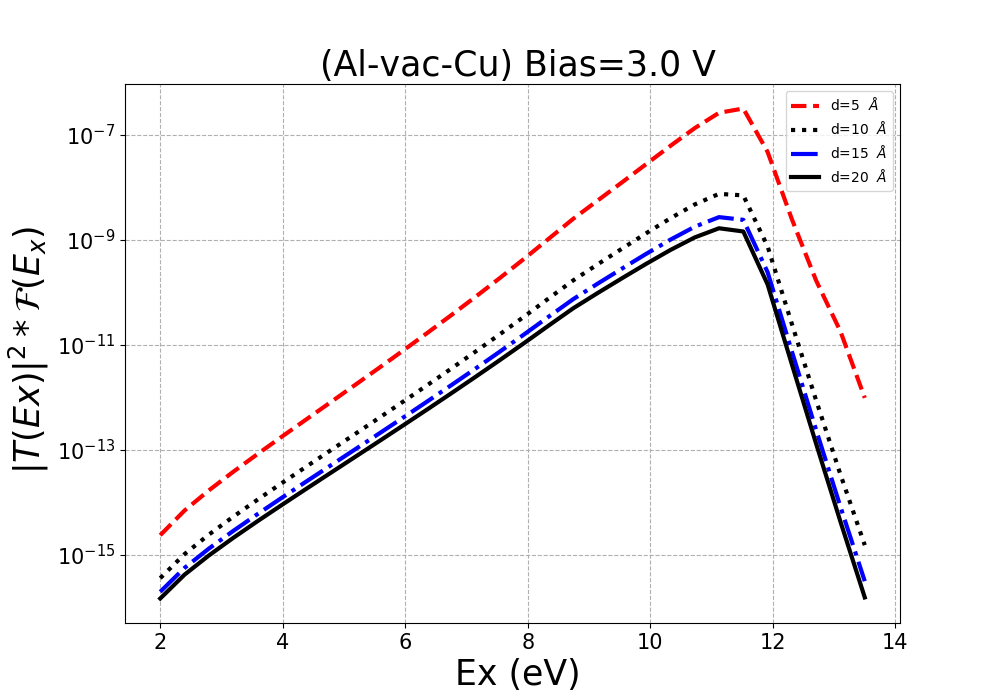}
	\caption{\,\,\,\,Plot \,\,\,\,of\, \,\,\,product\, \,\,of \,\,\,\,Transmission Probability\,\, \,and \,\,Fermi \,Factor \,versus\,\, $E_x$ \,for Bias Voltage = 3 V and $d=5,\,10,\,15,\,20$ $\AA$ }
	\label{fig:Figure7}	
\end{figure}

 Fig.\ref{fig:Figure7}, shows a plot of the product of calculated (Airy) $\rvert T(E_x) \rvert^2$ and the Fermi Factor $\mathcal{F}$ for metal electrodes Al(tip) and Cu(sample). The product $\rvert T(E_x) \rvert^2 \times \mathcal{F}$ increases exponentially upto Fermi level $(\eta_1 = 11.7\, eV)$ of the left electrode and drops sharply thereafter, well before reaching the upper limit of Energy Stage II $(\eta_1+\phi_1=15.86\, eV)$. Thus significant contribution to the tunneling current density arises from a narrow energy band (for $E_x$) around $\eta_1$. The Fermi Factor is thus seen to damp out the contribution to the tunneling current density for larger values of the kinetic energy $E_x$ for the translational motion of the electrons in the x-direction.
\begin{figure}[h]
	\centering
	\includegraphics[width=3.5in,height=2.5in]{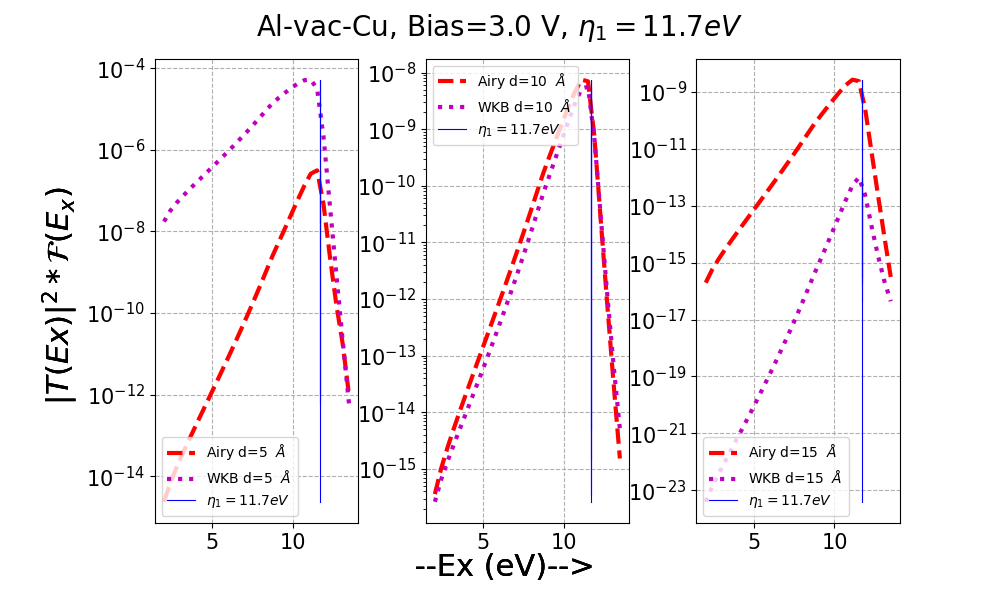}
	\caption{Plot of Transmission Probability ( Airy compared\,\, with \,\,WKB ) \,\,\,versus\,\,\, $E_x$\,\ for\,\, \,Bias Voltage = 3 V and $d=5,10$ $\AA$ (Al-vac-Cu).}
	\label{fig:WKB}
\end{figure}
\par \noindent Fig.\ref{fig:WKB} compares $\rvert T(E_x) \rvert^2 \times \mathcal{F}$ calculated using Airy functions with that using WKB approximation for an intermediate bias of $3$ V. For this bias voltage, this product $\rvert T(E_x) \rvert^2\times \mathcal{F}$ using the WKB approximation is found to exceed the Airy function determined $\rvert T(E_x) \rvert^2\times \mathcal{F}$ for low tip-sample distances such as $d = 5 \AA$. The two are nearly equal for $d = 10 \AA$, however its Airy value exceeds its WKB value for $d = 15 \AA$.
 Thus the WKB approximation does not provide an accurate  dependence of the tunnel current densities on the tip-sample distances. This feature suggests that determination of surface topogrophies of samples would be inaccurate, if the image processing software uses WKB approximation.
\subsection{\label{sec:level4}C. ${\bm {T(E_x)}}$ FOR VERY LOW BIAS}
\par In the low bias regime (0.1V to 1V) the Airy Function $Ai$ becomes very small causing underflow problems and the Airy function $Bi$ becomes very large causing overflow problems as in the calculation. Even introduction of denominators as in equation (\ref{phden19}), doesn't seem to help. Instead it is more useful to convert the Schr\"odinger equation to the following inhomogeneous equation
\begin{equation}\label{Greq}
\mathscr{L}\psi(x)=f(x)\psi(x)
\end{equation}  
where 
\begin{equation}\label{op}
\mathscr{L}=\dfrac{d^2}{dx^2}-\kappa^2
\end{equation} 
\begin{equation}\label{kap}
\dfrac{\hbar^2\, \kappa^2}{2m}=\eta_1+\phi_1-E_x
\end{equation}
\begin{equation}\label{fx}
\dfrac{\hbar^2}{2m}f(x) = -(\phi_1-\phi_2+eV_b)\dfrac{x}{d}
\end{equation}
The term $f(x)$ in the equation (\ref{Greq}), is usually very small for low bias voltages and low contact potentials, permitting a perturbative solution of equation (\ref{Greq}).
\par Fig.\ref{fig:Figure7}, shows that the maximum contribution to the tunneling occurs for $E_x$ lying in a very narrow energy band whose upper limit is $\eta_1$. For $E_x \leqslant \eta_1$ the minimum value of $\frac{\hbar^2 \, \kappa^2}{2m}$ will be $\phi_1$. The condition that $\frac{\hbar^2}{2m} |f(x)|$ be smaller than this minimum value of $\frac{\hbar^2 \, \kappa^2}{2m}$ is $eV_b < \phi_2$. Hence for low bias voltages, $|f(x)|< \kappa$ is true and perturbative solutions are valid. For $eV_b > \phi_2$ the Airy function methods will do fine. 
\par Define the Green function for the operator $\mathscr{L}$ of equation (\ref{op}) as 
\begin{equation}\label{Gr}
\mathscr{L}G(x,x')=\delta(x-x')
\end{equation}
Let the wavefunction in the barrier region be
\begin{equation}\label{21}
\psi(x) = \phi_0(x) + \phi(x) 
\end{equation}
\noindent where $\phi_0(x)$ is the solution to the homogeneous equation  
\begin{equation}\label{hom}
\mathscr{L}\phi_0(x)=0
\end{equation}
From equations (\ref{21}) and (\ref{hom}), $\phi(x)$ will satisfy 
\begin{equation}\label{23}
\mathscr{L}\phi(x)=f(x)\,[\phi_0(x)+\phi(x)]
\end{equation}
\begin{align}\label{ept}\nonumber
\phi(x')&=\int_{0}^{d}G(x,x')f(x)[\phi_0(x)+\phi(x)]\,dx \\
&+\, \text{end point terms}
\end{align}
The end point terms ($viz$ at $x=0$ and $x=d$) can be made to vanish if the Green function and the function $\phi$ are both made to vanish at  $x=0$ and $x=d$. Equation (\ref{ept}) can be iterated to obtain terms containing higher and higher powers of $f(x)$.  Here $f(x)$ will be regarded as small enough to neglect higher than first order terms. Keeping only the first order term, gives 
\begin{equation}
\phi(x')=\int_{0}^{d}G(x,x')f(x)\,\phi_0(x)dx 
\end{equation}
and therefore 
\begin{equation}\label{psi}
\psi(x)=\phi_0(x)+\int_{0}^{d}G(x,x')f(x')\,\phi_0(x')dx'
\end{equation}
\noindent The tunneling amplitude $T^{Gr}$ is calculated using the wavefunction in equation (\ref{psi}) and joining it smoothly with the wavefunctions given in equations (\ref{psi1}) and (\ref{psi3}). 
\par For very low bias, the energy of the electrons in the Energy Stage II is well above the Fermi level of the first electrode and the Fermi Factor is extremely small for these erergies. Therefore these electrons will make a negligible contribution to tunneling and hence the Green function method outlined above is not extended to the second stage. Instead, the second stage contribution for very low bias voltages is zeroed out.
\subsection{\label{sec:level3}D. CURRENT FROM CURVED TIPS USING PLANAR MODEL CURRENT DENSITIES}
As mentioned in the introduction it is necessary to go beyond the planar model and treat the tip as a sharp pointed surface and let the sample be flat. For a sharp tip, its radius of curvature at the point closest to the sample must be small, so that the length of the electric field lines increases rapidly as one goes away from the centre. Since the tunneling current densities fall off exponentially with increasing electric field line length therefore these current densities when integrated over the tip area will lead to a finite value for the current.  
\par
\begin{figure}[h]
	\centering
	\includegraphics[width=2.0in,height=2.1in]{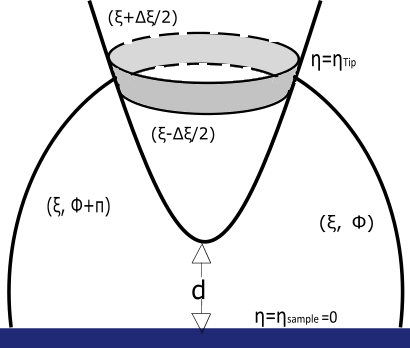}
	\caption{Hyperboloid Tip and Spheroidal shaped field line.} 
	\label{fig:tiparea}	
\end{figure}

 A good model in which the tip is sharp and the sample is flat is provided by using confocal hyperboloids for the tip and sample shapes (Fig.\ref{fig:tiparea}). These surfaces are coordinate surfaces in the prolate spheroidal coordinate system\cite{cutler}\cite{Patil8}\cite{date}\cite{russel}\cite{SG}\cite{jun}\cite{moon}\cite{morse}\cite{Aizpurua}. This is described by 
\begin{equation}
x=\rho \cos\phi, \quad y=\rho \sin\phi, \quad  z=a\xi \eta
\end{equation}
where $\rho=a \sqrt{\xi^2-1} \sqrt{1-\eta^2}$,  and $a$ is a constant whose value is determined by the tip radius of curvature $R$ and the tip sample distance $d$ as $a=\sqrt{d(d+R)}$.
The range of coordinate values are given by
\begin{equation} 
1\leqslant \xi \leqslant \infty, \, \, -1\leqslant \eta \leqslant 1, \, \,  0\leqslant \phi \leqslant 2\pi
\end{equation}
The Laplace equation for the electrostatic potential in this coordinate system becomes one dimensional and equipotential surfaces are confocal hyperboloids which are a one parameter family of surfaces, whose equation in cartesian coordinates is 
\begin{equation}\label{hyp}
\frac{z^2}{a^2\eta^2} - \frac{(x^2+y^2)}{a^2(1-\eta^2)}=1
\end{equation}
The tip surface (treated here as an emitter) is represented by a hyperboloid described by equation (\ref{hyp}) in which $\eta=\eta_{tip}=d/a$
The sample surface is characterised by  $\eta=\eta_{sample}=0$ which is the $x-y$ plane.
The equation of the field line that passes through the point $(\xi,\phi)$ is independent of $\phi$ and is given by
\begin{equation}
\frac{\rho^2}{a^2(\xi^2-1)}+\frac{z^2}{a^2\xi^2}=1
\end{equation} 
and the length of this field line stretching from $\eta =\eta_{sample}$ to $ \eta=\eta_{tip}$  is 
\begin{equation}
d_{fl}(\xi) = a \int_{\eta_{sample}}^{\eta_{tip}}\frac{\sqrt{\xi^2-\eta^2}}{\sqrt{1-\eta^2}} d\eta
\end{equation}
\par In Fig.\ref{fig:tiparea}, a small shaded area on the tip is shown along with, a couple of electric field lines, both lines, parametrized by $\xi$. This small shaded area on the tip, and the corresponding area on the sample define a planar tunnel junction of infinitesimal area, in which the length of the field line $d_{fl}(\xi)$ plays the role of the tip sample distance. Thus the net current density $J_{Net}(\xi)$ for this infinitesimal area planar tunnel junction is given by equation (\ref{jnet}). The product of the area of the shaded region $dS(\xi)$ with the current density of this infinitesimal tunnel junction gives the current carried by it. The sum of these currents for all such junctions covering a reasonably wide area on the tip will give the total net current for the entire non planar tip. Thus planar model current densities can be used to calculate the total net current for a given non planar shape of the tip. This procedure has been well described by Saenz and Garcia \cite{SG}.
\subsection{\label{sec:level6}THE CALCULATION }
The total current density $J_{Net}(\xi)$ is obtained by integrating the product of the tunneling probability and the Fermi Factor over the energy $E_x$ of the tunneling electrons for the translational motion along the x-direction. The tunneling amplitude $T=T[E_x,d_{fl}(\xi)]$ depends not only upon the energy $E_x$ but also on the location of the point on the tip (dictated by $\xi$) from where tunneling occurs. 
The current density $J_{Net}(\xi)$, given by an equation similar to equation (\ref{jnet}) is a function of $\xi$ through its dependence on the length of the field line $d_{fl}(\xi)$, and since $J(\xi)$ decreases rapidly with $d_{fl}(\xi)$ it will also decrease equally rapidly with $\xi$. The range of energies $E_x$ in the integration spans those in Energy Stages I and II described earlier in section B. 
\begin{figure}[h]
	\centering
	\includegraphics[width=3.5in,height=2.5in]{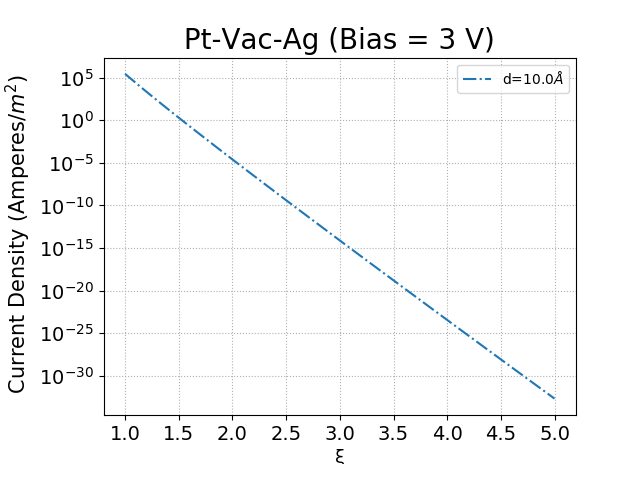}
	\caption{Plot of Current Density versus Length of field line with $R=20 A^{0}$, $d=10 A^{0}$ and Bias $=3$ V for Pt-vac-Ag} 
	\label{fig:xi}	
\end{figure}
\par The net current is given by 
\begin{equation}\label{curr}
 I= \int_{1}^{\xi_{max}} J(\xi)\, dS(\xi) 
\end{equation}
where the upper limit $\xi_{max}$ is chosen to have a value of $5$, because the current density $J_{Net}(\xi)$ is essentially zero beyond this value as shown in Fig.\ref{fig:xi}. In this figure it can be seen that the current density drops by roughly $20$ orders of magnitude as $\xi$ increases from $\xi=1$ to $\xi=5$. The surface element $dS(\xi)$ in equation (\ref{curr}) is given by 
\begin{equation}
 dS(\xi)=2\pi a^2 \sqrt{(\xi^2-\eta_{tip}^2)\,\, (1-\eta_{tip}^2)}\,d\xi
\end{equation}
is the area of the shaded region on the tip as shown in Fig.\ref{fig:tiparea}. The curved line in this Fig.\ref{fig:tiparea} is the field line correponding to the parameter $\xi$ $(1 \leqslant \xi \leqslant \xi_{max})$. 
\subsection{\label{sec:level7}RESULTS AND DISCUSSION }
\par
\begin{figure}[h]		
	\centering                 
	\includegraphics[width=3.5in,height=2.5in]{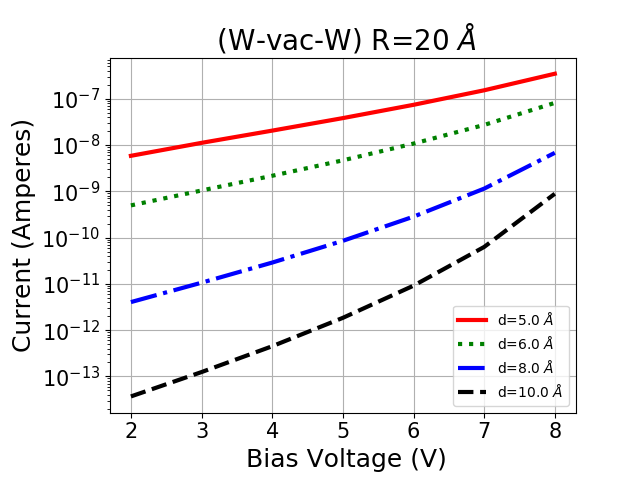}		
	\caption{Plot of Current versus Bias Potential for $R=20\,\, A^{0}$ and $d=5,\, 6,\, 8,\, 10\,\,  A^{0}$ \,\,for similar electrodes}
	\label{fig:2a}
\end{figure}

 Figure \ref{fig:2a} shows calculated I-V characteristics for identical electrodes (W-vac-W) in which the tip is negatively biased, for several tip-sample distances. The current increases almost exponentially with bias voltage at all distances.
\begin{figure}[h]
	\centering		
	\includegraphics[width=3.5in,height=2.5in]{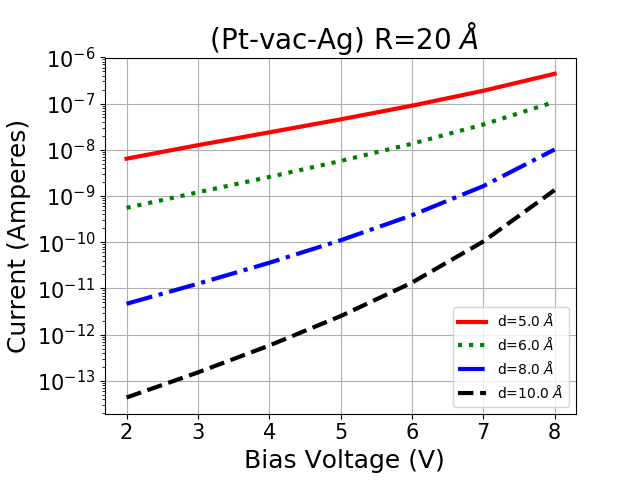}		
	\caption{Plot of Current versus Bias Potential for $R=20\,\, A^{0}$ and $d=5,\, 6,\, 8,\, 10\,\,  A^{0}$ \,\,for dissimilar electrodes}
	\label{fig:2b}
\end{figure}
Figure \ref{fig:2b} shows calculated I-V characteristics for dissimilar electrodes (Pt-vac-Ag) and the I-V curves behave similar to the case of identical electrodes, $viz$ the current increases almost exponentially with bias voltage for all distances.
\begin{figure}[h]
	\centering
	\includegraphics[width=3.5in,height=2.5in]{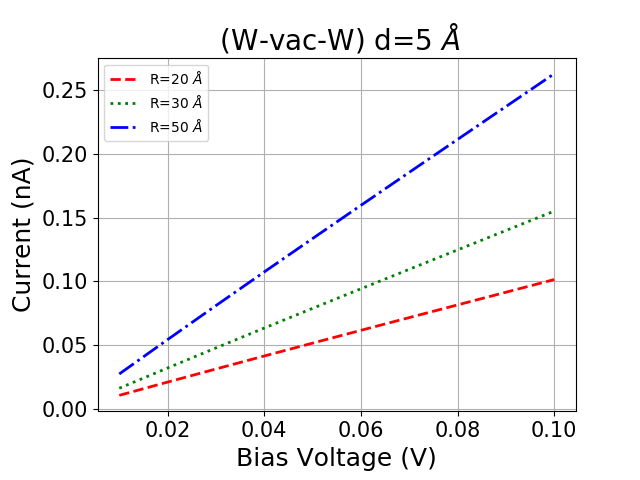}
	\caption{Plot of Current versus Bias Potential $(< 0.1\, V)$ for $R=20, 30, 50 A^{0}$ and $d=5 A^{0}$ for W-vac-W} 
	\label{fig:Figure7a}	
\end{figure}  
\par Ting {\it et. al.}\cite{ting} have reported values of tunneling currents in the W-vac-W electrode system for very low bias voltages and for several cross currents $I_\text{cross}$, including for $I_\text{cross}=0$. Here $I_\text{cross}$ is a current made to flow in the surface of the sample. For $I_\text{cross}=0$, their experimental setup corresponds to the tip-vacuum-sample system for which tunneling currents are calculated in this paper. Tunneling currents of $0.1 \text{nA}$ are reported by Ting {\it et. al.}\cite{ting} for low bias voltages in the range $[0 \,\text{to} \,70\, \text{mV}]$. Fig.\ref{fig:Figure7a} plots the calculated tunneling current for (W-vac-W) for very low bias for a fixed tip sample distance $d = 5 \,\AA$ and several tip radii of curvature $(R=20, 30, 50\,\AA)$. These plots show the tunneling current increasing with the bias voltage as expected. Also the tunneling currents are seen to increase with increasing tip radii. Further for $R=30$ $\AA$, a good numerical agreement with results for $I_\text{cross}=0$ of Ting {\it et. al.}\cite{ting} is obtained. 
\begin{figure}[h]
		\centering
		\includegraphics[width=3.5in,height=2.5in]{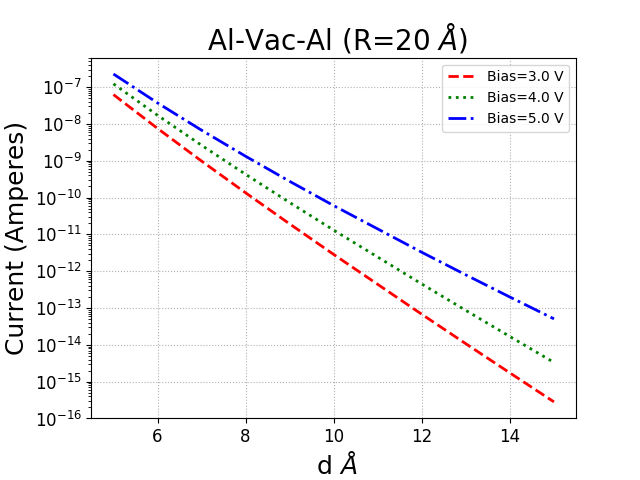}
		\caption{Plot \,of \,Current versus \,Tip - Sample distance for $ R=20  A^{0}$ and Bias Voltage = 3, 4 and 5 V for similar electrodes}
		\label{fig:2c}
	\end{figure}
	\begin{figure}[h]
		\centering
		\includegraphics[width=3.5in,height=2.5in]{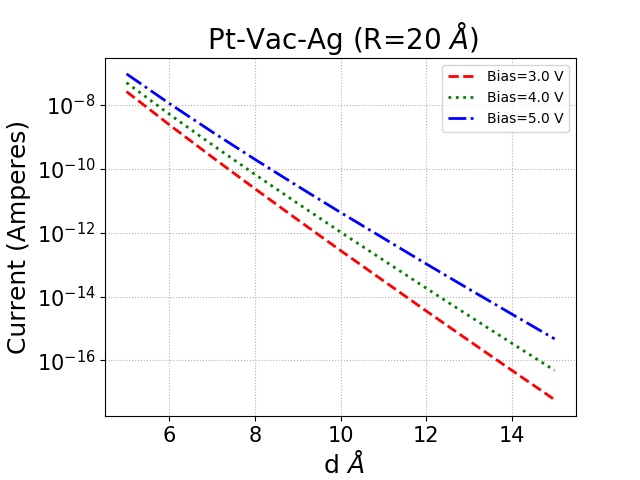}		
		\caption{Plot \,of \,Current versus \,Tip - Sample distance for $ R=20  A^{0}$ and Bias Voltage = 3, 4 and 5 V for Dissimilar electrodes}
		\label{fig:2d}
	\end{figure}
\par Figures \ref{fig:2c} and \ref{fig:2d} show the plots of calculated tunneling currents $I$ versus tip sample distance $d$ in similar (Al-vac-Al) and dissimilar (Pt-Vac-Ag) pairs of electrodes. These plots are for bias voltages of $3$ V, $4$ V and $5$ V, and tip radius of curvature of $R=20$ $\AA$. Therefore $I$ is found to decrease exponentially with increasing $d$. This  behaviour is also qualitatively reproduced by almost all calculations of tunneling current density including those that use the WKB approximation \cite{simmonsI}\cite{simmonsII}\cite{SBPZ}\cite{Hofer}. 
\par
\begin{figure}[h]
	\centering
	\includegraphics[width=3.5in,height=2.0in]{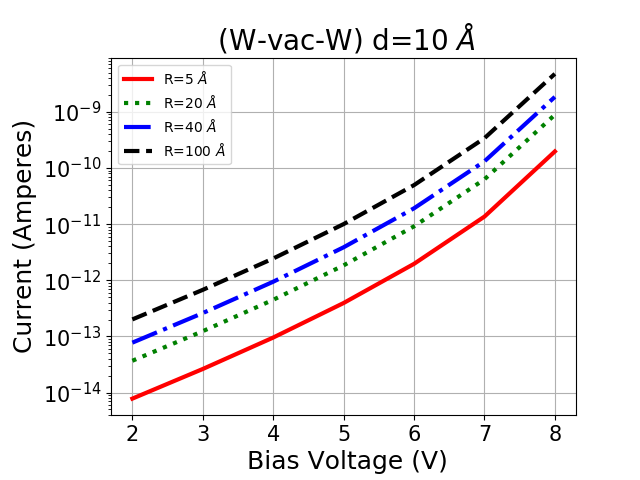}			
	\caption{Plot of Current versus Bias Voltage for $d=10$ $A^{0}$ and $R=5,20,40,100\, A^{0}$}
	\label{fig:Fig3}
\end{figure}

 For fixed bias voltage $V$ and tip sample distance $d$, increasing the tip radius $R$, leads to a flatter tip, which means that more area in the vicinity of the center lies closer to the sample and the current density for these areas is greatly increased. Since the current $I$ is an integral of the current density over tip area, $I$ is expected to increase with increasing $R$ for all $V$ and $d$. Fig.\ref{fig:Fig3} indeed confirms this behaviour by showing how the $I-V$ curves  correspond to higher currents for larger $R$. The general behaviour of $I$ verses $V$ is not much influenced by $R$.
\begin{figure}[h]	
	\centering
	\includegraphics[width=3.5in,height=2.0in]{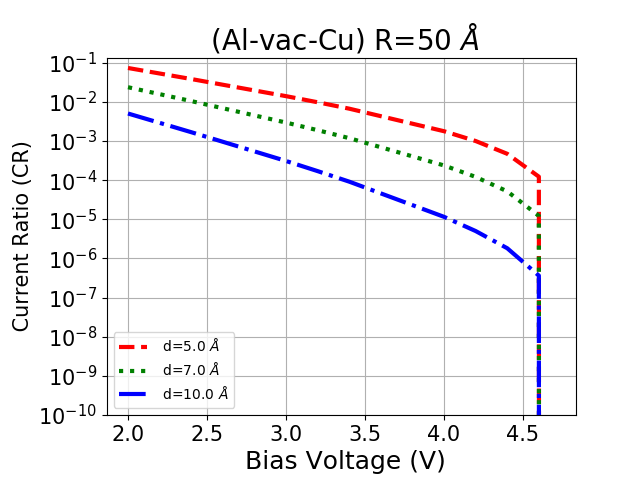}
	\caption{Log Plot of Current Ratio (CR) as a function of Bias Voltage ($2$V to $5$V) \,\, for $R=50\AA$ and $d=5, 7, 10\AA$.}
	\label{fig:CR1}		
\end{figure} 
\begin{figure}[h]	
	\centering
	\includegraphics[width=3.5in,height=2.0in]{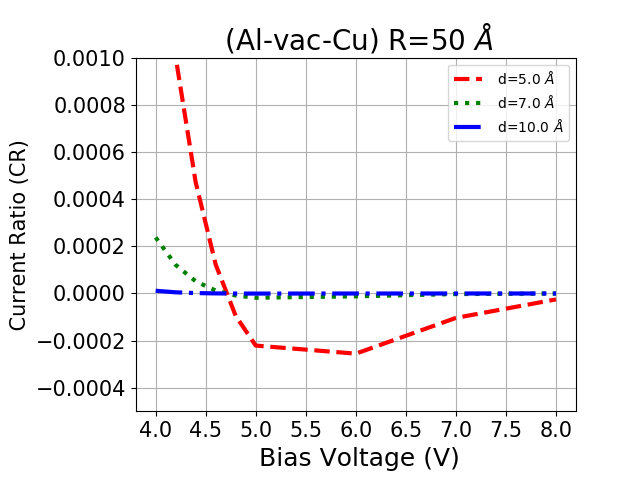}
	\caption{Plot of Current Ratio (CR) as a function of Bias Voltage ($4$V to $8$V) for $R=50\AA$ and $d=5,7,10\AA$	}
	\label{fig:CR2}		
\end{figure} 
The Pauli Effects have been discussed in some detail following Figures \ref{fig:FF},\ref{fig:ThTerm1},and \ref{fig:ThTerm2}. The effect of the third term on the tunneling currents is studied by considering the Current Ratio $\text{CR}= 1-(I_\text{Non-Pauli}/I)$. Figures \ref{fig:CR1},\ref{fig:CR2} show that the ratio of the third term contribution to the net tunneling current which is labelled as the CR is quite small for all bias voltages. The value of the CR and hence the strength of the Pauli Effect decreases very fast with increasing tip-sample distance. The CR is positive for bias voltages less than $eV_b=\eta_1-\eta_2=4.7$eV, for which voltage range $I_\text{Net} > I_\text{Non-Pauli}$ and for bias voltages greater than this value the CR is negative, implying that $I_\text{Net} < I_\text{Non-Pauli}$. The Pauli Effect has a larger magnitude at low bias voltages as compared to higher bias voltages. 
\begin{figure}[h]
	\centering
	\includegraphics[width=3.5in,height=2.0in]{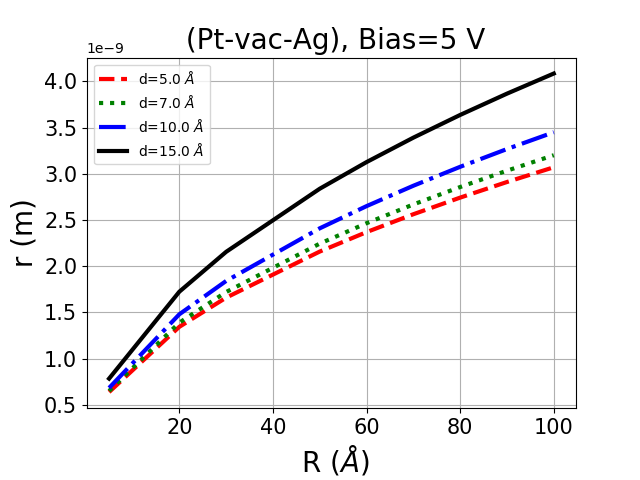}
	\caption{\,Plot \, of \, r \, ( Measure \, of \, lateral Resolution ) versus R ( \,Radius of \,curvature of the tip ) for\, \, \, Bias Voltage = 5 V \,and\, Tip sample distances = 5, 7, 10, 15 $\AA$}
	\label{fig:Figure8}	
\end{figure}
\par The ratio $\dfrac{I}{J_{Net}(\xi=1)}$ gives an area $\sigma$, which can be regarded as the area of a circle of lateral resolution; {\it i.e.}, sample profile features that lie within this circle would be poorly resolved. The radius $r=\sqrt{\sigma/\pi}$ of this circle would therefore be a good measure of the lateral resolution limit, and can be named as the Lateral Resolution Parameter (LRP).
\begin{figure}[h]
	\centering
	\includegraphics[width=3.3in,height=2.0in]{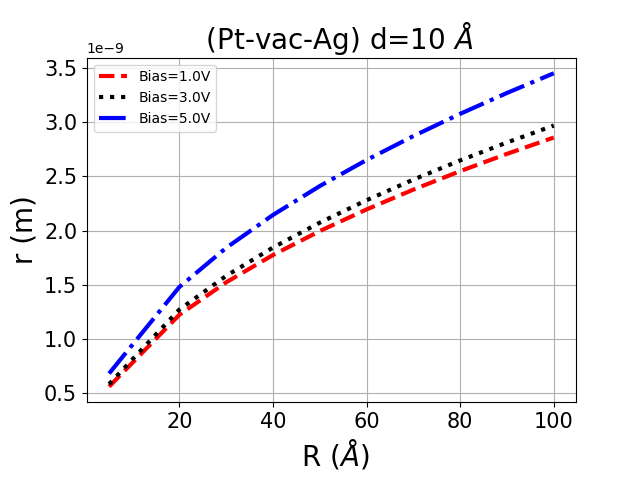}
	\caption{\,Plot \, of \, r \, ( Measure \, of \, lateral Resolution ) versus R ( \,Radius of \,curvature of the tip ) for\, \, Tip sample distances = 10 $\AA$ and Bias Voltage=1, 3, 5 V
	}
	\label{fig:Figure9}
\end{figure} 
\par Fig.\ref{fig:Figure8} and Fig.\ref{fig:Figure9} show the dependence of the $r$ (LRP) with the tip radius $R$  which is varied from $(5 \,\AA\, \text{to}\, 100 \, \AA)$. Both figures show that sharper the tip ($viz$ lower the value of $R$), the greater is the lateral resolution ($viz$ smaller the value of $r$). Fig.\ref{fig:Figure8} shows that for fixed bias the lateral resolution is degraded with increasing tip sample distance. In Fig.\ref{fig:Figure9} the tip sample distance is kept constant, while the bias voltage is changed. This figure shows that the lateral resolution is degraded as the bias voltage increases; which feature is found to be more pronounced at higher, rather than at lower bias voltages. 
\subsection{\label{sec:level8}SUMMARY}
The model used for calculating tunnelling current for the biased M-V-M tunnel junctions treats the effective barrier potential as a trapezoidal potential. Exact analytic expressions for the tunneling probability and current are obtained by solving the Schr\"odinger’s equation for the trapezoidal potential, for which the wavefunctions can be expressed in terms of Airy functions. Electrons with x-directed translational energies ranging from $0$ eV to $\eta_1+\phi_1$ contribute to tunneling.  
\par The current is calculated for a wide range of input parameters such as tip-sample distance, bias voltage, and radius of curvature of the tip, both electrodes being at room temperature. Pauli Effects are explicitly introduced and their behaviour as a function of bias voltage and tip-sample distance is studied. A measure of lateral resolution of the tunnel junction is introduced and its behaviour as a function of tip radius of curvature for several values of bias voltage, tip sample distance is also presented. 

\end{document}